\documentclass[prl,10pt,floatfix,twocolumn,superscriptaddress,amsmath,amssymb,showpacs]{revtex4-1}
\usepackage[pdftex]{graphicx}
\usepackage{dcolumn}
\usepackage{bm}
\usepackage{amssymb}
\usepackage{comment}

\begin{document}

\title{Predicting plasticity with soft vibrational modes: from dislocations to glasses} 
\author{J\"org Rottler} \date{\today}
\affiliation{Department of Physics and Astronomy, University of British Columbia, 6224 Agricultural Road, Vancouver, BC V6T1Z4, Canada}
\author{Samuel S. Schoenholz}
\affiliation{Department of Physics, University of Pennsylvania, Philadelphia, Pennsylvania 19130, USA}
\author{Andrea J. Liu}
\affiliation{Department of Physics, University of Pennsylvania, Philadelphia, Pennsylvania 19130, USA}

\begin{abstract}
We show that quasi localized low-frequency modes in the vibrational spectrum can be used to construct soft spots, or regions vulnerable to rearrangement, which serve as a universal tool for the identification of flow defects in solids. We show that soft spots not only encode spatial information, via their location, but also directional information, via directors for particles within each soft spot.  Single crystals with isolated dislocations exhibit low-frequency phonon modes that localize at the core, and their polarization pattern predicts the motion of atoms during elementary dislocation glide in exquisite detail.  Even in polycrystals and disordered solids, we find that the directors associated with particles in soft spots are highly correlated with the direction of particle displacements in rearrangements.
\end{abstract}

\maketitle 

Plastic flow of crystalline solids is controlled by the dynamics of structural defects\cite{cottrell_mechanical_1964,schmid_plasticity_1968}.  In crystals perfect enough that isolated dislocation lines can be identified, a powerful topological framework exists to predict their response to applied shear stress \cite{hirth_theory_1992}. Real crystalline materials, however, can have dislocations that organize along multiple grain boundaries.  Once individual dislocations can no longer be resolved, predicting where the solid will fail locally becomes much more difficult.  

The extreme limit of a polycrystalline material is an amorphous solid, where the notion of topological defects is entirely lost. 
However, anharmonic properties, such as the energy barriers that need to be overcome in plastic flow, are strongly correlated with harmonic properties, such as the vibrational mode frequency~\cite{xu_anharmonic_2010}.   Similarly, rearrangements (anharmonic phenomena) tend to occur in regions with lower local shear modulus (a harmonic property)~\cite{yoshimoto_mechanical_2004,tsamados_local_2009}.  As a result, the vibrational normal modes of inherent structures\cite{widmer-cooper_localized_2009,brito_geometric_2009,tanguy_vibrational_2010,manning_vibrational_2011,chen_measurement_2011} contain information about the localized rearrangements that occur in amorphous solids\cite{schall_structural_2007,falk_dynamics_1998}. In particular, the quasi localized low-frequency vibrational modes identify a population of flow defects or soft spots in disordered packings\cite{manning_vibrational_2011}. Thus, these modes tell us \emph{where} local failure tends to occur.

\begin{figure}[t] 
\begin{centering} 
\includegraphics[width=8cm]{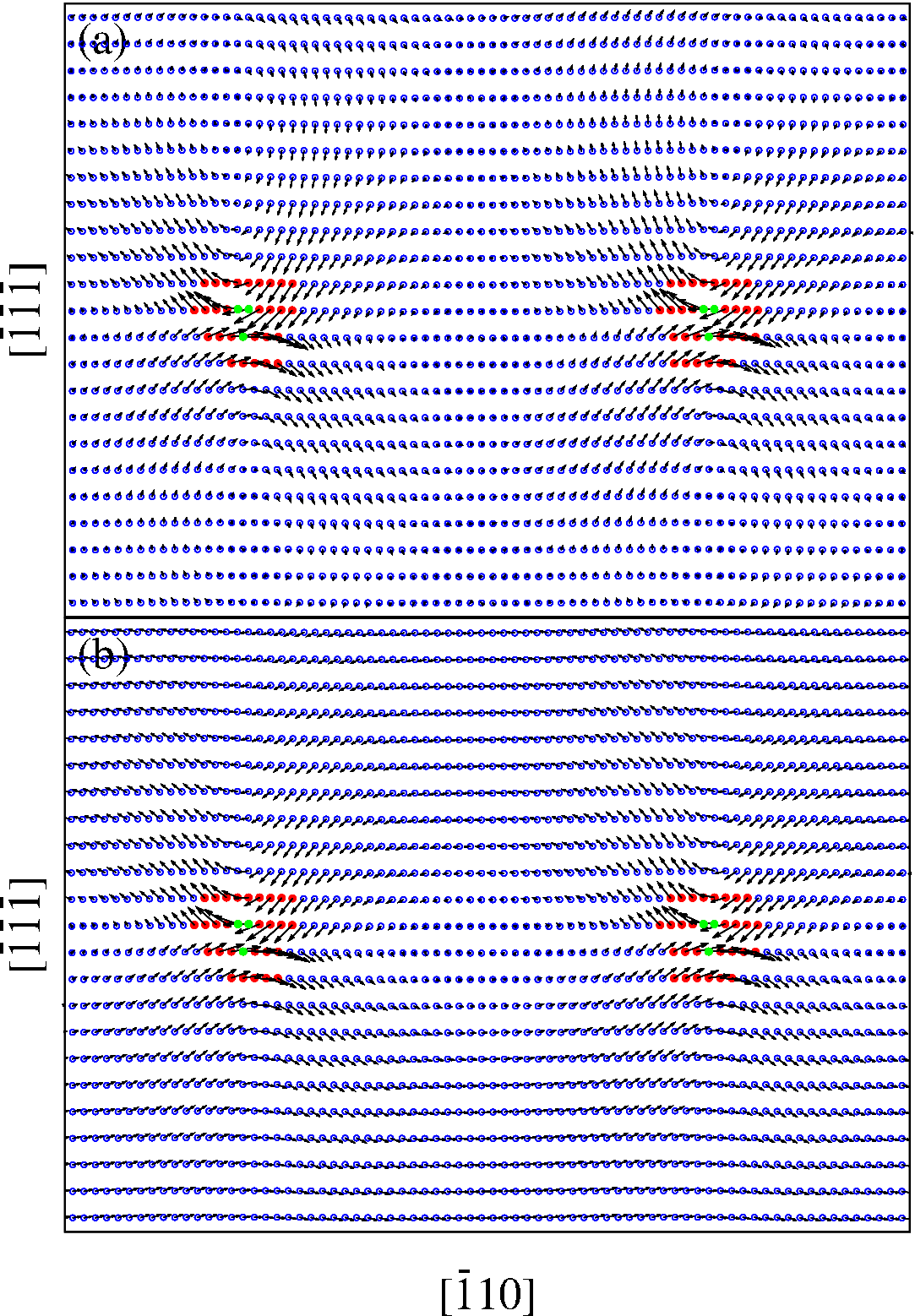}
\caption{\label{3d-fig}(Color online) Projection onto the  [${\bar 1}\bar{1}$2]-direction of 3D split partial dislocations connected by a stacking fault in a fcc lattice composed of 38,000 particles. (a) Polarization field of the first localized mode at dislocation core (green), and soft spot obtained from the 500 largest polarization vectors (red). Particles with less or more than 12 neighbors are shown in green. (b) Displacement field resulting from an elementary dislocation slip.}
\end{centering}
\end{figure}

In this letter, we show that low-frequency quasi localized modes also tell us \emph{how} the particles will move during local failure, in systems ranging from crystals to fully disordered solids.   For each particle in a soft spot, we introduce a director based on the orientations of the polarization vectors of the particle from all the modes that contribute to that particle.   As a result, each soft spot not only contains spatial information in its location, but also contains directional information for each particle within it. We find that the direction of a particle's displacement during a rearrangement correlates strongly with its director for all materials regardless of their degree of disorder.  

We begin by introducing a single edge dislocation along the $[\bar{1}10]$-direction into a perfect fcc crystal. In all our calculations, particles interact with a 6-12 Lennard Jones potential that is truncated but interpolated so that both first and second derivatives of the potential vanish smoothly.  The potential energy is minimized using a combination of conjugate gradient and damped dynamics methods (FIRE algorithm) \cite{bitzek_structural_2006}.  In close packed structures, edge dislocations can lower their energy by splitting into two Shockley partial dislocations connected by a stacking fault according to the dislocation reaction $[{\bar 1}10]/2\rightarrow [\bar{2}11]/6+[\bar{1}2\bar{1}]/6$.  In Fig.~\ref{3d-fig}(a) we show the atomic configuration near the cores (green) of the two partial dislocations, which can be identified as particles that have 11 or 13 nearest neighbors, respectively. The crystal is nonperiodic in the  [${\bar 1}\bar{1}\bar{1}$]-direction in order to avoid the creation of dislocation dipoles, and atomic positions are projected along the [${\bar 1}\bar{1}$2] line direction (into plane) as the configuration still has cylindrical symmetry. Among the lowest energy vibrational eigenmodes, we find one mode that localizes at the core of the two partials. Its polarization field is also shown in Fig.~\ref{3d-fig}(a); the 500 particles with the largest polarization vector magnitudes are highlighted (red). These particles all cluster around the dislocation core (green), confirming that it is the center of a soft spot.

Edge dislocations are highly mobile and contribute to plastic deformation by gliding in the direction of their Burgers vector. We produce this elementary slip event by shearing the top and bottom plane of atoms parallel to the $[\bar{1}10]$-direction and reminimizing the total energy.  Both partials glide simultaneously as they overcome the Peierls barrier and their cores move by one lattice constant; the displacement field is shown in Fig.~\ref{3d-fig}(b). Remarkably, the displacements near the core in Fig.~\ref{3d-fig}(b) agree almost perfectly with the polarization field of the soft mode shown in Fig.~\ref{3d-fig}(a).  Equivalent results were obtained for a single edge dislocation in a 2D triangular lattice.  Thus, for isolated edge dislocations, a single vibrational mode contains all information required to predict the particle trajectories associated with elementary dislocation glide. 

\begin{figure}[t] 
\includegraphics[width=7cm]{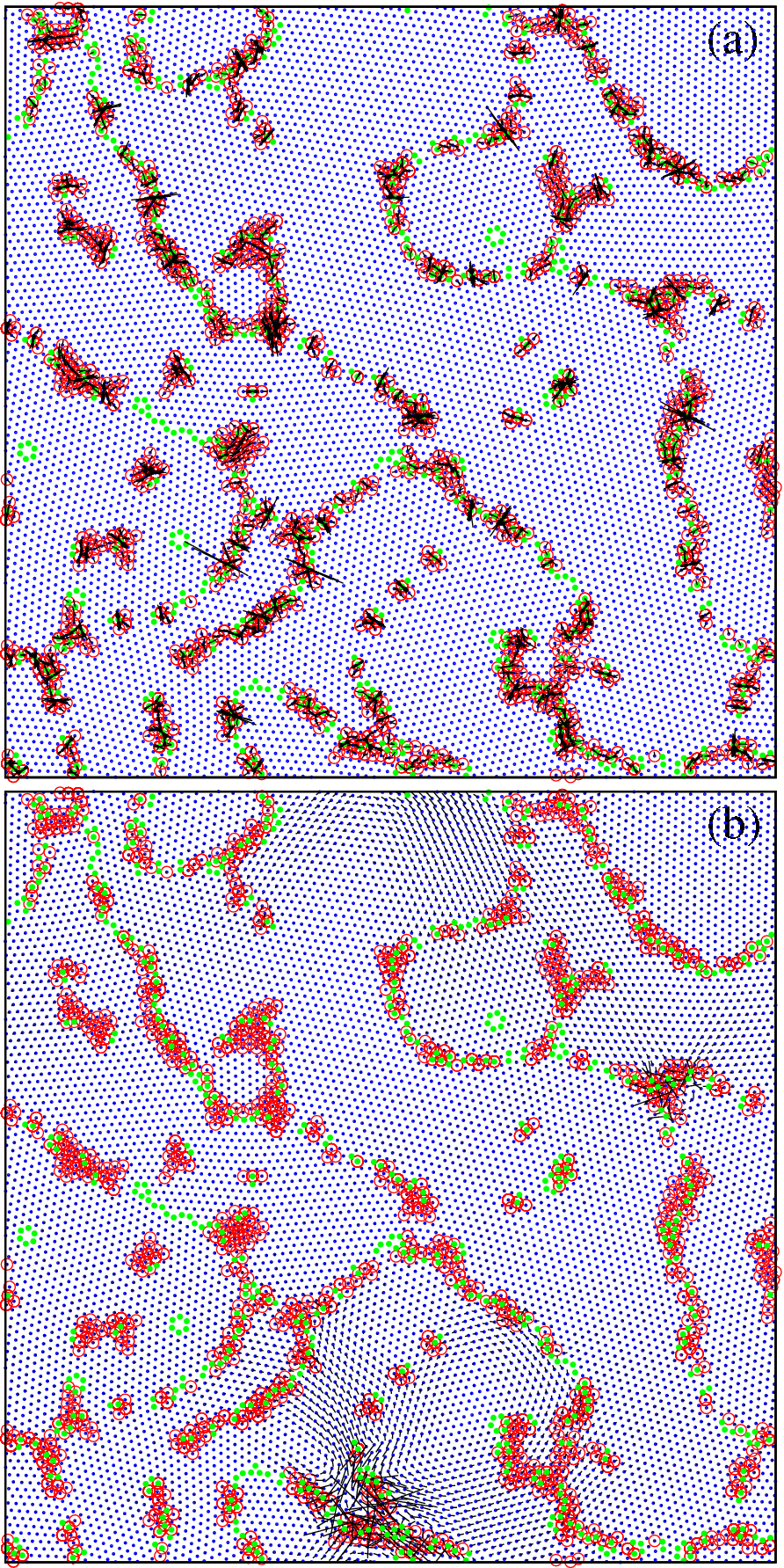}
\caption{\label{poly-fig}(Color online) 10,000 particle polycrystal in 2D. (a) Example configuration with grain boundaries (green), soft spots (red circles), and the polarization directions of the modes that define the soft spots (black lines). (b) Nonaffine displacement map after step shear strain of $10^{-3}$ (black lines).}
\end{figure}

We now depart from the idealized limit of isolated defects and explore increasingly disordered systems. First, we consider nanocrystalline samples in two dimensions (see Fig.~\ref{poly-fig}(a), generated by randomly placing atoms within the simulation box and then quenching the configuration to zero temperature by minimizing the energy). The system contains multiple grains of varying crystallographic orientations separated by grain boundaries (GB). Topological analysis easily locates the GBs as atoms that do not have 6 nearest neighbors; this analysis also identifies several vacancies. 

Following ref.~\cite{manning_vibrational_2011}, we build a soft spot population from $N_p$ particles with the largest polarization vector magnitudes in the $N_m$ lowest energy modes. Many particles appear in multiple modes. To focus specifically on quasilocalized low energy modes, we rank modes with polarization field $\{{\bf u}_i\}$ by their participation ratio $pr=(\sum_i {\bf u}_i^2)^2/(N \sum_i {\bf u_i}^4)$.  It is clear that the choice of $N_m$ and $N_p$ determines the soft spot covering fraction. Here we choose $N_m=300$ and $N_p=30$, which optimizes the difference between overlap with local plastic activity (see below for definition and supplemental material for details) and the covering fraction. We furthermore refine the soft spot population by discarding small isolated clusters of size less than four~\cite{manning_vibrational_2011}. This choice can be justified from inspection of the soft spot size distribution, which shows an excess number of small clusters that are unlikely to be related to plastic activity, but in fact, it makes almost no difference to the quantitative results. 

Fig.~\ref{poly-fig}(a) shows the resulting population of soft spots (red circles). Soft spots are found exclusively at GBs, in agreement with recent observations in colloidal crystals \cite{chen_phonons_2013}, and have a covering fraction of about 9\%.  Note that vacancies are {\emph not} soft spots, as they do not contribute to plasticity~\cite{chen_phonons_2013}. 

In addition to particles in the soft spots, we superimpose in Fig.~\ref{poly-fig}(a) the polarization directions for all modes that contribute to a given particle. These directions are not randomly oriented, but have strong local correlations. A particle in a soft spot is therefore endowed with a preferred direction.  We define an orientation for each particle in a soft spot, as follows.  Because the forward and backward polarization directions are equivalent in harmonic theory, we calculate the total orientation tensor $U_{ij}=\langle \hat{u}_i\hat{u}_j-\delta_{ij}/2 \rangle$ for each particle in the soft spot, where $\hat {u}_i$ is $i$th component of the polarization vector for that particle in a given mode that contributes to the soft spot particle, i.e. if the mode is one of the $N_m$ lowest frequency modes {\emph and} the particle is one of the $N_p$ particles with the highest polarization vector magnitude for that mode. 

We now subject the material to a sizable step shear strain of $10^{-3}$, of order the typical strain between rearrangements, and reminimize the potential energy. The nonaffine displacements of all particles that result from this deformation can be seen in Fig.~\ref{poly-fig}(b) together with the soft spots and GBs. In this example, two plastic events have occurred near triple junctions of two GBs. The largest nonaffine displacement vectors indicate regions of maximal plastic activity, and these regions overlap well with the soft spot population. Moreover, the nonaffine displacements seem to point preferentially along the directions of the soft modes. 

\begin{figure}[t] 
\begin{centering} 
\includegraphics[width=7.5cm]{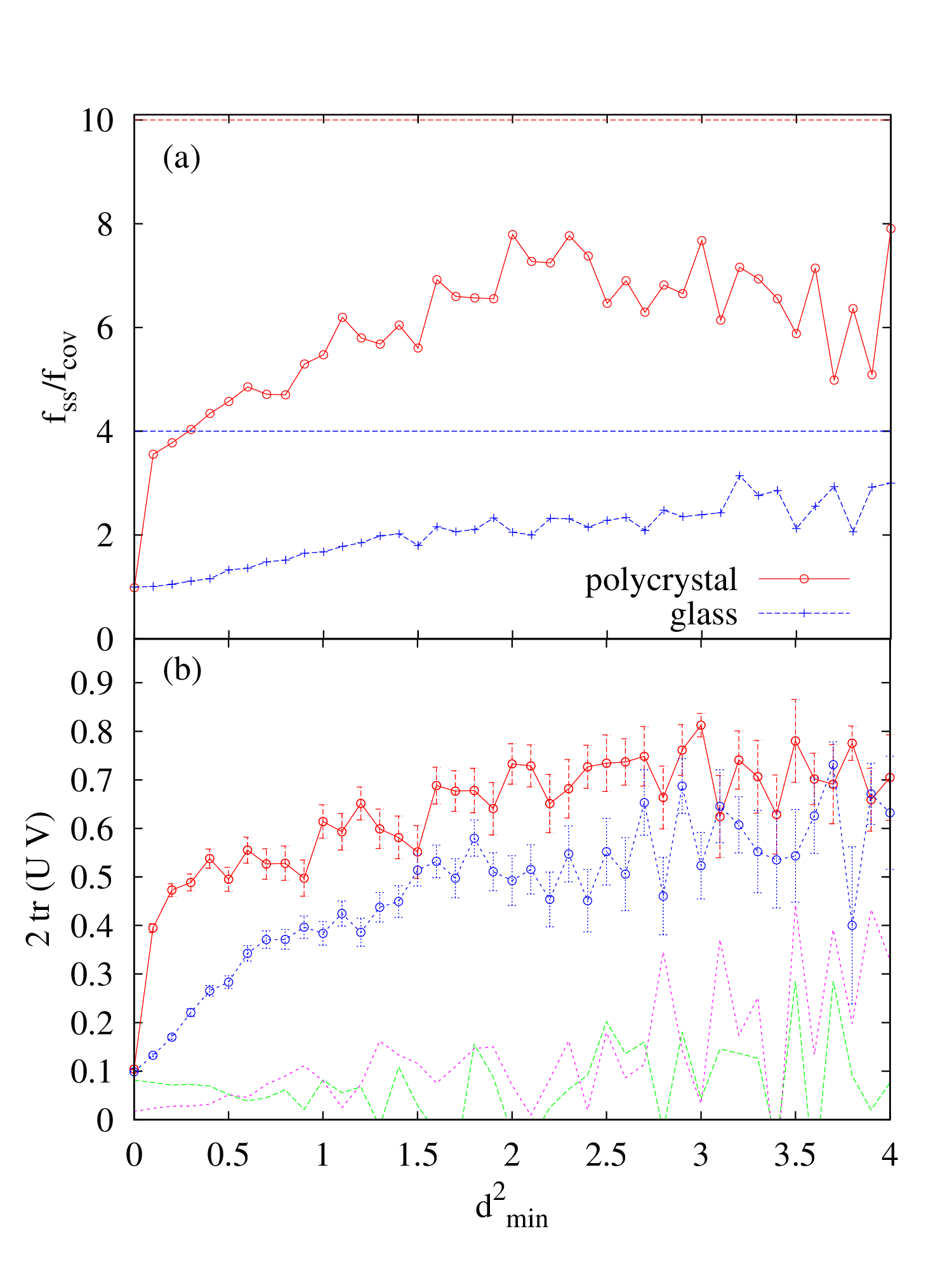}
\caption{\label{corr-fig}(a)(Color online) Fraction of particles in a soft spot as
  a function of $d^2_{\text{min}}$, divided by the soft spot covering fraction. Data represent an average over 1000 different
  configurations subjected to forward/backward simple
  shear. (b) Correlation of the total orientation $U_{ij}$
  with the displacement tensor
  $V_{ij}$ (see text), as a function of $d^2_{\rm min}$. Dashed (dotted) lines show for comparison the alignment of plastic displacements with only the lowest energy (low participation ratio) mode in the glassy system.}
\end{centering}
\end{figure}

To quantify soft spot - plasticity correlations we average over 1000 different GB configurations subjected to step shear strains of $\pm 10^{-3}$, as in Fig.~\ref{poly-fig}(b).  For some of these configurations, we observe avalanches that cover most of the system~\cite{bailey_avalanche_2007,lemaitre_rate-dependent_2009,karmakar_statistical_2010,rodney_modeling_2011,salerno_avalanches_2012}. Since we are interested in predicting the {\emph onset} of failure rather than the entire avalanche, we remove these realizations from the statistical average by removing values of $d^2_{\text{min}}>6\, \sigma^2$. (If we retain them, the correlation between soft spots and rearrangements is slightly reduced but there is no qualitative difference; see supplemental material).  As a scalar measure of local plastic activity we use the minimum mean squared difference $d_{\text{min}}^2$ of displacements of a group of atoms relative to a central one and those they would have if the strain were uniform \cite{falk_dynamics_1998}.  Figure \ref{corr-fig}(a) shows the fraction of atoms that are found in a soft spot as a function of their $d_{min}^2$ value, normalized by the soft spot covering fraction. In the absence of plastic activity ($d_{\text{min}}^2=0$), this ratio is indeed close to one, indicating no correlations. For nonzero values of $d_{\text{min}}^2$, however, the ratio rises rapidly and saturates at a value near 8 for $d_{min}^2\ge 2$. A plastic event is thus 8 times more likely to occur at a soft spot than if the soft spots were uncorrelated with rearrangements.

To quantify the degree of alignment between soft spot directors and the direction of displacements during a rearrangement,  we calculate the trace of the product, $2\,{\rm tr}(U\cdot V)$, of the orientation tensors $U_{ij}$ for all particles in the soft spot and $V_{ij}=\hat{v}_i\hat{v}_j-\delta_{ij}/2$, where $\hat{v}$ is the normalized nonaffine displacement vector of the particle during a rearrangement event. The average is taken over all particles in the soft spots and over all configurations studied. This correlation function is one if $\hat{u}$ and $\hat{v}$ are perfectly parallel or antiparallel and zero on average if their relative orientation is random. 

In Fig.~\ref{corr-fig}(b) we show the value of this correlation function averaged over all particles that contribute to a soft spot as a function of the amount of plastic activity as indicated by $d_{min}^2$. The correlation is small at low values of  $d_{min}^2$, but rises rapidly with increasing plastic deformation and saturates near 0.7 for $d_{\text{min}}^2>3$, corresponding to displacements of order the particle diameter (rearrangements). The modes corresponding to soft spots no longer predict the pattern of atomic motion perfectly as we saw for isolated dislocations, but still predict displacement directions to a remarkable degree.  

\begin{figure}[t] 
\includegraphics[width=7cm]{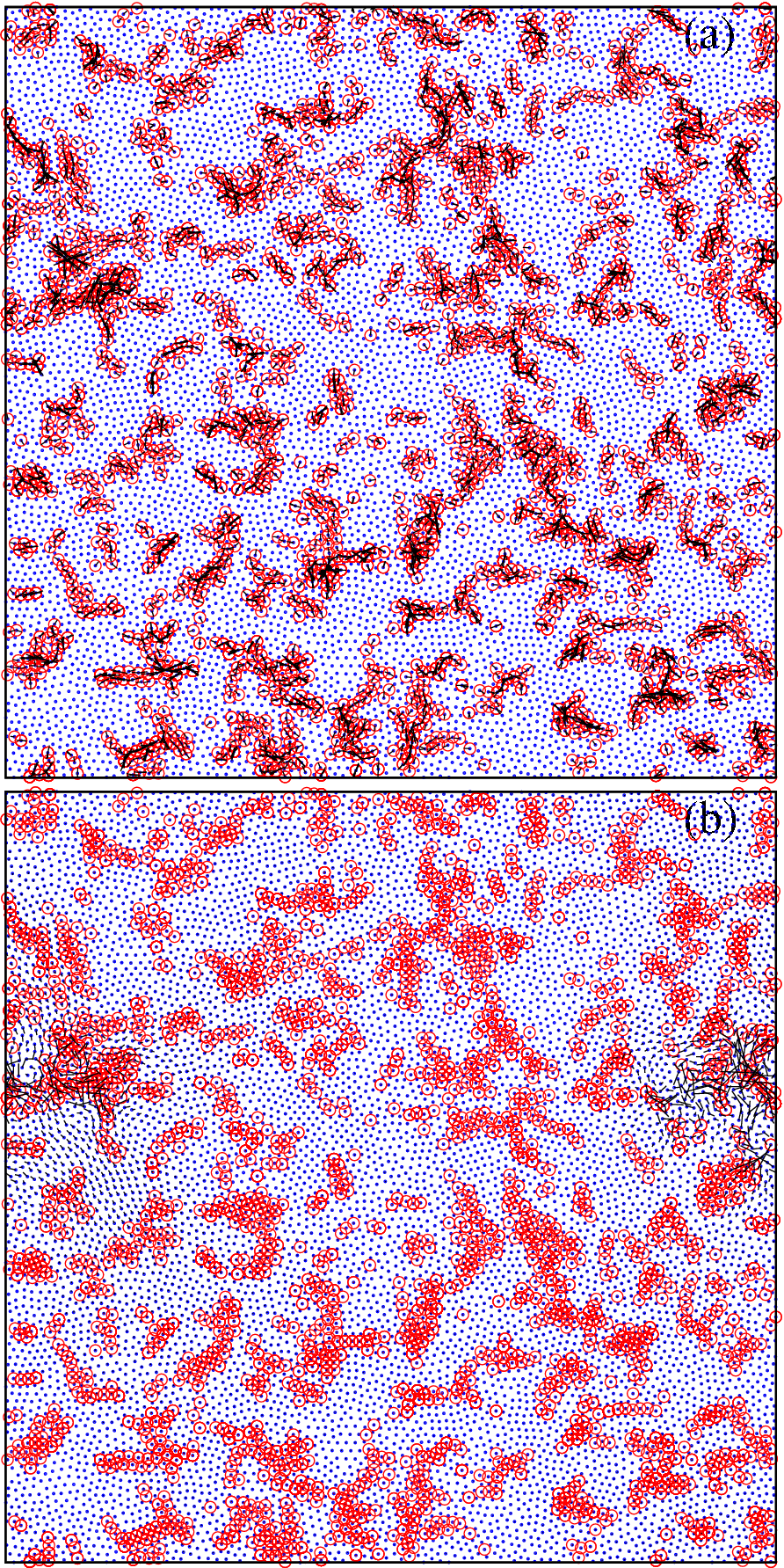}
\caption{\label{glass-fig}(Color online) 10,000 particle amorphous solid in 2D. (a) Example configuration with soft spots (red) and polarization directions from the contributing soft modes (black lines). (b) Soft spot map with nonaffine displacements from a step shear strain of $10^{-3}$ (black lines).}
\end{figure}

The ultimate application of the soft spot concept lies in amorphous solids. We obtain fully disordered, two-dimensional packing from a binary 65-35 Lennard-Jones mixture of small and large particles with well characterized glass forming properties \cite{bruning_glass_2009}.  Particles are placed randomly within the simulation box and their positions are quenched to zero temperature via energy minimization. In the configuration shown in Fig.\ref{glass-fig}(a), we can no longer identify topological defects, but we construct a population of soft spots with the same approach described for polycrystals. As optimal parameters we determine $N_m=250$, $Np=30$, yielding an average soft spot covering fraction of 23\%. As in the polycrystal case, we superimpose on each atom in a soft spot the polarization directions of the soft modes in order to visualize the preferred directions. 

The response of an amorphous packing to a step shear strain of $10^{-3}$ can be seen in Fig.~\ref{glass-fig}(b). The nonaffine displacement map reveals a somewhat extended event with a vortex-like pattern. In this example, the plastic event includes multiple soft spots with larger displacement amplitudes closer to the soft spots. In order to quantify the ability of the soft spots to predict plastic deformation, we again consider 1000 different realizations of the disorder, and again discard those deformations that generate nonlocal avalanches. In Fig.~\ref{corr-fig}(a) we also plot the probability to observe a particle in a soft spot as a function of the local plastic amplitude $d_{min}^2$. The probability rises from one to about three with increasing $d_{min}^2$. The correlation with plastic activity is thus weaker than in the polycrystal, but a plastic rearrangement is still about three times more probable at a soft spot. Fig.~\ref{corr-fig}(b) shows the degree of alignment between nonaffine displacement and soft mode polarization as measured from the orientation parameter. Remarkably, the degree of alignment saturates at a value above 0.5 that is only slightly lower than for the polycrystal.

We emphasize that in the disordered system, 12 modes on average contribute to a soft spot so that no one mode dominates the director for a given particle in a soft spot. To demonstrate this point directly, we also show in Fig.~\ref{corr-fig}(b) the alignment between the plastic displacement field and the polarization vectors for the lowest energy mode or the lowest participation ratio mode at low frequency. The correlations are negligible, showing that the strong correlation between the directors of particles in soft spots and their displacements in rearrangements are not a trivial result of the fact that a single quasi localized mode approaches zero frequency at a rearrangement in a system under quasi static shear.  

We have shown that low-frequency quasi localized modes are a powerful tool for identifying flow defects that can be applied seamlessly in solids ranging from crystals to fully disordered glasses.  In all cases, the flow defects, or soft spots, not only predict {\emph where} rearrangements are likely to occur but also {\emph how} the particles displace during rearrangement. This correlation weakens somewhat as the amount of disorder increases and the topologically distinct features of dislocations disappear.  However, it remains quite strong even in the amorphous state, where the soft spots remain as echoes of structural defects.  

Low-frequency quasi localized modes arise because sound waves scatter from disorder. Our findings therefore suggest that in all solids, the type of disorder that is most effective in scattering sound waves is flow defects.  Thus, there is an intimate connection between the scattering of sound modes and the carriers of plasticity in solids, whether crystalline or amorphous.

JR thanks C.~W.~Sinclair for instructive discussions and acknowledges a UBC Killam Faculty Research Fellowship for financial support.  This work was primarily supported by the UPENN MRSEC, NSF-DMR-1120901 (JR, SS and AJL).

%

\end{document}